\documentclass[12pt,tightenlines,eqsecnum,floats,aps,amsmath,amssymb,nofootinbib,prd,shownopacs]{revtex4}
\usepackage{amsmath,amssymb,amsfonts,amsthm}
\usepackage{graphicx}
\usepackage{enumerate} 
\usepackage{colordvi} 

\usepackage[mathscr]{eucal}
\usepackage[latin1]{inputenc}
\usepackage{amsmath}
\usepackage{amsfonts}
\usepackage{amssymb}
\usepackage{graphicx}

\def\lp{{\ell}_{\rm Pl}}

\def\q{\mathring{q}}
\def\e{\mathring{e}}
\def\w{\mathring{\omega}}

\newcommand{\rcr}{\rho_{\mathrm{crit}}}

\newcommand{\f}{\frac}

\def\rmax{\rho_{\mathrm{max}}}

\def\rcr{\rmax}

\usepackage{enumerate}

\def\f{\frac}

\def\d{\textrm{d}}

\def\mb{\bar \mu}

\usepackage{enumerate}

\newcommand{\be}{\nopagebreak[3]\begin{equation}}
\newcommand{\ee}{\end{equation}}
\newcommand{\ba}{\nopagebreak[3]\begin{eqnarray}}
\newcommand{\ea}{\end{eqnarray}}

\newcommand{\bmult}{\nopagebreak[3]\begin{multline}}
\newcommand{\emult}{\end{multline}}

\def\lp{{\ell}_{\rm Pl}}

\def\f{\frac}

\def\d{\textrm{d}}

\def\mb{\bar \mu}

\def\rmax{\rho_{\rm max}}
\def\smax{\sigma^2_{\rm max}}
\def\sina{\sin{(\bar{\mu}_1 c_1)}}
\def\sinamu{\frac{\sin{(\bar{\mu}_1 c_1)}}{\bar{\mu}_1}}
\def\sinb{\sin{(\bar{\mu}_2 c_2)}}
\def\sinbmu{\frac{\sin{(\bar{\mu}_2 c_2)}}{\bar{\mu}_2}}
\def\sinc{\sin{(\bar{\mu}_3 c_3)}}

\def\cosa{\cos{(\bar{\mu}_1 c_1)}}
\def\cosb{\cos{(\bar{\mu}_2 c_2)}}
\def\cosc{\cos{(\bar{\mu}_3 c_3)}}
\def\xa{\sqrt{\frac{p_2 p_3}{p^3_1}}}
\def\xb{\sqrt{\frac{p_1 p_3}{p^3_2}}}
\def\xc{\sqrt{\frac{p_1 p_2}{p^3_3}}}

\def\xbxc{\frac{p_1}{p_2 p_3}}

\def\Heff{\mathcal{H}_{\rm eff}}
\def\Hmatt{\mathcal{H}_{\rm matt}}
\def\mua{\bar{\mu}_1}
\def\mub{\bar{\mu}_2}
\def\muc{\bar{\mu}_3}
\def\lp{l_{\rm Pl}}

\def\d{{\rm d}}

\usepackage{enumerate}
\begin{document}

\title{Contrasting features of anisotropic loop quantum cosmologies: the role of spatial curvature}

\author{Brajesh Gupt}
\email{bgupt1@lsu.edu}

\author{Parampreet Singh}
\email{psingh@phys.lsu.edu}
\affiliation{Department of Physics and Astronomy, Louisiana State University,
Baton Rouge, Louisiana 70803, USA}


\begin{abstract}
A characteristic feature of  loop quantization of the isotropic and Bianchi-I spacetimes is the existence of universal bounds on the energy density and the expansion and shear scalars, independent of the matter content. We investigate the properties of these physical quantities in Bianchi-II and
Bianchi-IX spacetimes, which have been recently loop quantized using the connection operator approach.
 Using the effective Hamiltonian approach, we  show that for Bianchi-II spacetime, energy density and the expansion and shear scalars turn out to be bounded, albeit not by universal values. In Bianchi-IX spacetime, when the approach to the classical singularity is isotropic, above physical quantities are bounded. In addition, for all other cases, where the approach to singularities is not isotropic and effective dynamics can be trusted, these quantities turn out to be finite. These results stand in sharp distinction to general relativity, where above physical quantities are generically unbounded, leading to the break down of geodesic equations. In contrast to the  isotropic and Bianchi-I models, we find  the role of energy conditions for Bianchi-II model and the inverse triad modifications for Bianchi-IX to be significant to obtain above bounds. These results bring out subtle physical distinctions between the quantization  using holonomies over closed loops performed for isotropic and Bianchi-I models, and the connection operator approach.
We find that qualitative differences in physics exist for these quantization methods even for the isotropic models in the presence of spatial curvature. We highlight these important differences in the behavior of the expansion scalar in the holonomy based quantization and connection operator approach for isotropic spatially closed and open models.
\end{abstract}

\maketitle


\section{Introduction}

Singularities in general relativity are primarily characterized by the divergences in the curvature invariants and the break down of geodesic evolution. A central piece in the singularity theorems of Hawking, Penrose and Geroch is the Raychaudhuri equation, which determines the evolution of the congruence of geodesics via the properties of the expansion $(\theta)$ and shear $(\sigma^2)$ scalars, and the components of the stress energy tensor. As the singularity is approached, these quantities blow up, resulting in the inextendability of geodesics. The  scalars $\theta$ and $\sigma^2$, also capture the extrinsic and the Weyl curvature of the spacetime, and hence turn out to be useful measures to gain insights on the behavior of curvature invariants and the nature of singularities.
 It is generally believed that a quantum theory of gravity will shed important insights on the resolution of singularities. If such a theory allows an
effective spacetime description to understand the behavior of geodesics, it is pertinent to ask whether $\theta$ and $\sigma^2$, along with the components of the stress-energy tensor are bounded by the quantum gravitational effects, and if yes, under what conditions.  Understanding of these properties is vital to gain insights on the generic resolution of singularities in quantum gravity and the underlying requirements for geodesic completeness.

In recent years, a lot of progress has been made in the quantization of homogeneous spacetimes in loop quantum cosmology (LQC) to pose such questions. LQC is a non-perturbative canonical quantization of homogeneous cosmologies, based on loop quantum gravity (LQG), which
predicts resolution of singularities in various situations \cite{as}. These include spatially flat $(k=0)$ isotropic model sourced with a massless scalar field \cite{aps1,aps2,aps3}, in presence of cosmological constant \cite{bp,ap,kp} and inflationary potential \cite{aps4}, spatially closed ($k = 1$) model \cite{apsv,warsaw_closed}, spatially open ($k = -1$) model \cite{kv,szulc}, Bianchi-I model \cite{awe2} and Bianchi-II \cite{awe3} and Bianchi-IX spacetimes with a massless scalar field \cite{we1}.  In all of these models, the quantum Hamiltonian constraint turns out to be non-singular, which is a direct consequence of the underlying quantum discreteness predicted by LQG.  Further, one recovers classical GR in the limit when the spacetime curvature becomes small.
Quantum evolution in LQC is  governed by a quantum difference equation, however, for  a class of semi-classical states, there exists an effective continuum spacetime description.  This allows one to obtain an effective Hamiltonian constraint for LQC \cite{jw,vt,psvt}.\footnote{The modified Einstein's equations, can also be obtained from an effective action in LQC using Palatini approach \cite{os}.}
The resulting Hamilton's equation lead to the modified Einstein's equations which turn out to describe the underlying quantum evolution extremely accurately \cite{aps1,aps2,aps3,bp,ap,kp,warsaw_closed}.\footnote{It should be noted, that for these numerical simulations, bounce occurs at volumes greater than the Planck volume.}  
The modified Einstein's equations have been widely used in conjunction with the analytical and numerical techniques to capture the details of the underlying physics in LQC and to reveal rich phenomenological features \cite{as}.

In LQC, one starts with a classical phase space in Ashtekar variables: the SU(2) connection $A^i_a$ and conjugate triads $E^a_i$, which are then symmetry reduced to $c^i$ and $p_i$ respectively, by incorporating the underlying symmetries of a homogeneous spacetime. The elementary variables used for quantization are the holonomies of the connection components and fluxes corresponding to the  triads.\footnote{These turn out to be proportional to triads in the homogeneous setting.} The Hamiltonian constraint, the only non-trivial constraint left after symmetry reduction, is then expressed in terms of these elementary variables and quantized. This procedure leads to two novel features. The first of these arises by expressing field strength of the connection in terms of holonomies over a closed loop shrunk to a minimum area on the quantum geometry. This leads to a non-local nature of the field strength, which results in a quantum difference equation. The second feature arises due to the presence of inverse triad (or inverse volume) operators. Since the eigenvalue spectrum of triad operator is discrete and includes zero as an eigenvalue, its inverse it not densely defined. Using a classical identity in the phase space, inverse triads are expressed in terms of the Poisson brackets between holonomies and the  positive powers of triads \cite{tt}, and then quantized. The resulting eigenvalues of such an  inverse triad operator show a significant departure from the classical behavior near $p_i = 0$. When the triad component vanish, the eigenvalue of such an operator is zero. At larger values compared to the Planck value, it approximates the classical behavior. However, such modifications can only be consistently defined when the spatial manifold is compact. In case it is non-compact, the modifications to the classical behavior depend on the fiducial volume of the fiducial cell introduced to define the symplectic structure. Since this cell is an infra-red regulator introduced to regulate infinities occurring due to the underlying non-compactness, physics must be independent of it. Indeed, in the limit when the fiducial volume of the fiducial cell becomes infinite for the non-compact models, the inverse triad modifications vanish. Thus, for spatial manifolds which are non-compact, quantum geometry leads to new physics only via non-local nature of the field strength. If the spatial manifold is compact, apart from the non-local field strength, inverse triad effects can also be important.

Let us now discuss  some of the main features of the isotropic and  Bianchi-I models in LQC. These are the following: (i) For all of the isotropic and the  Bianchi-I models, the field strength can be expressed using holonomies, which are almost periodic functions of the connection $A^i_a$, over a closed loop. Quantum geometry fixes the minimum area of the loop using the eigenvalues of the area operator in LQG. It results in {\it a universal bound} on the energy density\footnote{This value coincides with the supremum of the expectation values of the energy density operator in the physical Hilbert space in an exactly soluble model in the spatially flat case \cite{acs}.}  ($\rcr \approx 0.41 \rho_{\mathrm{Planck}}$) and the expansion scalar in these models, independent of the choice of matter (and hence the energy conditions). These bounds have been shown to result in a generic resolution of all strong curvature singularities in isotorpic and spatially flat LQC \cite{ps09}. An analysis of various singularities, including the various exotic ones, strongly indicates this result to extendable to spatially curved models \cite{sv}.  The non-local nature of above field strength is also responsible for a universal bound on directional Hubble rates and $\sigma^2$ in the Bianchi-I model, which has been demonstrated to yield resolution of all strong singularities for different types of matter \cite{ps11}; (ii) Inverse triad corrections are potentially significant for only those universes which attain a size comparable to Planck value. If bounce of the universe occurs when its volume is much bigger than Planck volume, then inverse triad corrections play little role. In such cases, they are found to be neither responsible for bounds on the energy density and $\theta$, nor do they lead to any significant effects on the modified dynamics \cite{aps3,apsv}.



This uniformity of results for isotropic models with different spatial curvatures and Bianchi-I model is noteworthy. We recall  that for the $k=1$ model, construction of the closed loop is technically challenging where one can not simply extend the strategy used in the $k=0$ model.\footnote{In the spatially flat model, construction of such a loop is straightforward due to the availability of commuting left invariant vector fields on the spatial manifold. In the spatially closed  model, left invariant vector fields do not commute, and the loop is constructed using both left and right invariant vector fields \cite{apsv}.} The resulting quantization leads to non-trivial terms in the quantum Hamiltonian constraint which capture the spatial curvature. In contrast, the quantization of $k=-1$ model follows neither the approach in $k=0$ model nor $k=1$ model to express the field strength in terms of holonomies. This is due to a technical difficulty resulting from the presence of the off-diagonal terms in the spin connection $\Gamma^i_a$. Currently available quantizations overcome this problem by considering holonomies of the extrinsic curvature \cite{kv,szulc}, and  a priori  an agreement on the detailed physics at the Planck scale is unexpected. Further, in the Bianchi-I model, freezing of the anisotropic degrees of freedom does not lead to the quantization of the isotropic flat model. Instead one must integrate out the anisotropic degrees of freedom \cite{awe2}.
 Despite various differences in the quantization strategy of these spacetimes and the resulting subtleties, it is rather remarkable that loop quantization of isotropic models and Bianchi-I spacetime, reveals the same  bounds on the physical quantities independent of the energy conditions and are also not affected by inverse triad modifications for spacetimes with non-vanishing spatial curvature.

The goal of this manuscript is to analyze the physics of Bianchi-II and Bianchi-IX spacetimes in LQC, in the above context using the effective Hamiltonian approach. In these models, due to the interplay of spatial curvature and anisotropies, it is not possible to express field strength in terms of holonomies which are almost periodic functions of the connection components. A new strategy is needed to loop quantize these spacetimes, which was proposed in Ref. \cite{awe2}. One expresses the field strength in terms of a non-local connection defined via holonomies over open segments.\footnote{In terms of the symmetry reduced connection, in this approach, the connection operator $\hat c$ is defined as: $\hat c = \widehat{\frac{\sin(\bar \mu c)}{\bar \mu}}$.} The underlying quantum geometry does not directly constrain the length of such open segments. However, by demanding that the resulting expression for the field strength operator agrees with the one in the Bianchi-I model, one can fix the minimum length of the edge over which the holonomy is computed using the minimum quantum of area in LQG. The pertinent question is whether this strategy leads to physics at Planck scale which confirms with that established by the quantization of all other models in LQC or does it lead to novel surprises? In particular, are energy density, $\theta$ and $\sigma^2$ bounded in Bianchi-II and Bianchi-IX models? Are these bounds universal?
 What are the contributions of the modifications originating from the inverse triad operators for the Bianchi-IX model?  Are these modifications important or are they  insignificant, as in the case for the isotropic models?

We will show in this manuscript that the connection operator approach leads to some unexpected results. We find that energy density and the expansion and shear scalars are bounded in Bianchi-II model, however these bounds are not universal as they depend on the energy conditions.\footnote{In Bianchi-II model, the energy density has been shown to be bounded for the case of a massless scalar field \cite{awe3}. Since the matter content was fixed, the role of energy conditions was not known.}  In the Bianchi-II model, one must assume that the energy density is bounded below, else the shear scalar  diverges in the Planck regime. In the Bianchi-IX model, inverse triad modifications turn out to be critical to obtain these bounds when the approach to classical singularity is isotropic, i.e. all triads approach the singularity at the same time.  For the other types of singularities, we show that energy density and the expansion and shear scalars are finite except for two cases. These two cases correspond to the vanishing of one or two of the three triads. At this stage, it is neither evident whether the effective dynamics can be trusted in such a regime nor if such solutions exist in the Bianchi-IX model for some matter content.  These results bring out contrasts between the connection operator approach \cite{awe3,we1} and the holonomy approach \cite{aps1,aps2,aps3,apsv,aps4,ck,szulc,awe2}. We find that  demanding the consistency of connection operator approach with Bianchi-I model, which has vanishing spatial curvature, does not guarantee its consistency for the spatially curved models. In particular, the isotropic limit of the effective Hamiltonian constraint in the loop quantization of Bianchi-IX model does not lead to the effective Hamiltonian constraint of the $k=1$ model \cite{apsv}, but to that of a different quantization -- the inequivalent connection operator quantization of $k=1$ model \cite{ck}, and that too only in the regime where inverse triad modifications can be ignored. Unlike the holonomy based quantization of $k=1$ model, where expansion scalar is bounded by a global maxima in the effective spacetime description, $\theta$ does not have a maximum in the connection operator approach. The same turns out to be true for an alternate quantization of $k=-1$ model based on the connection operator approach.

The organization of this paper is as follows. We start with a summary of the effective Hamiltonian in LQC for Bianchi-I model and revisit the analysis in Ref. \cite{csv} to show that the energy density, and the expansion and the shear scalars are bounded by universal values for arbitrary matter content. Here we obtain the correct bound on the shear scalar, which was previously estimated incorrectly. In Sec. 3, we repeat the analysis for non-compact Bianchi-II model. Here we show that the energy density, the expansion and the shear scalar are bounded only if one imposes energy conditions. We show that if one allows energy density to be unbounded below, the shear scalar diverges. In Sec. 4, we analyze the effective Hamiltonian constraint for the Bianchi-IX model and show that inverse triad corrections play an important role for $\rho, \theta$ and $\sigma^2$  to be bounded for the isotropic approach to singularities. Due to the inverse triad modifications, there is an additional subtlety in the behavior of energy density, which is also addressed. Here we also discuss the isotropic limit of the effective Hamiltonian in Bianchi-IX model, and find that it does not lead to the one for $k=1$ loop quantization \cite{apsv}. In the absence of inverse triad modifications  the limit is given by an alternative connection operator based quantization, which has expansion scalar unbounded. Whereas in the presence of inverse triad modifications, the isotropic limit agrees with  none of the available isotropic quantizations. We summarize the results with a discussion in Sec. 5. In Appendix A, we discuss the behavior of expansion scalar in $k=1$ model in the holonomy based and the connection operator quantizations in LQC. A similar discussion for  $k=-1$ model is provided in Appendix B. These show that for isotropic $k=\pm 1$ models, the expansion scalar is unbounded for the effective Hamiltonian constraint corresponding to the connection operator approach.

\section{Bianchi-I model}
Bianchi-I model is one of the simplest examples of an anisotropic spacetime. It has vanishing spatial curvature and in the isotropic limit it yields the $k=0$ Robertson-Walker (RW) metric. The homogeneous Bianchi-I anisotropic spacetime can be described by a manifold $\Sigma \times \mathbb{R}$ where $\Sigma$ is the 3-spatial hypersurface characterized by a set of three commuting Killing vectors. As earlier works in LQC, we will consider the topology of $\Sigma$ as $\mathbb{R}^3$. The spacetime metric of Bianchi-I spacetime is given by,
\begin{equation}
ds^2=-N^2d\tau^2+a_1^2dx^2+a_2^2dy^2+a_3^2dz^2
\end{equation}
where $a_1$,$a_2$ and $a_3$ are the directional scale factors. These can be used to define a mean scale factor $a := (a_1 a_2 a_3)^{1/3}$. Since the spatial manifold is non-compact, in order to define the symplectic structure, one needs to introduce a fiducial cell ${\cal V}$. This cell's edges can be chosen to lie along the integral curves of fiducial triads $\e^a_i$ and have coordinate length $l_1, l_2, l_3$.  The cell ${\cal V}$ has fiducial volume $V_o=l_1 l_2 l_3$ with respect to the fiducial metric $\q_{ab}$ compatible with fiducial co-triads $\w^i_a$.

In LQG the canonical variables are the Ashtekar connection $A_a^i$ and the triad $E^a_i$ which, which due to the symmetry of Bianchi-I spacetime, can be expressed in terms of components $c^i$ and $p_i$ as:
\be
A_a^i \, = \, c^i \, (l_i)^{-1} \, \w_a^i, ~~ \mathrm{and} ~~ E_i^a \, = \, p_i \, l_i\, V_o^{-1}\, \sqrt{\q}
\ee
The connection and triad components, $c^i$ and $p_i$, satisfy the following Poisson bracket relation,
\begin{equation}
  \{c^i,p_j\}=8\pi G\gamma \delta^i_{j}
\end{equation}
where $\gamma \approx 0.2375$ is Barbero-Immirzi parameter. The triads $p_i$ are related to the directional scale factors as
\begin{equation}
p_1 \, =\, \varepsilon_1 \, l_2\, l_3\, |a_2 \, a_3|, ~~ p_2 \, =\, \varepsilon_2\, l_1 \, l_3 \, |a_1 \, a_3|, ~~ p_3 \, = \, \varepsilon_3 \, l_2 l_1 |a_1 a_2|
\end{equation}
where $\varepsilon_i = \pm 1$ depending on the orientation of the triads. Without any loss of generality, we will choose the orientation to be positive in the following analysis.

Let us first consider the Hamiltonian constraint in the classical theory. For lapse $N = V$, the classical Hamiltonian constraint in terms of Ashtekar variables can be written as,
\begin{equation}
\mathcal{H}_{\mathrm{cl}}=\frac{1}{8 \pi G \gamma^2} \, \left(c_1 p_1 c_2 p_2 + \mbox{cyclic terms}\right) +\Hmatt V~,
\end{equation}
where $\mathcal{H}_{\mathrm{matt}}$ is the matter Hamiltonian and $V$ denotes the physical volume of the cell ${\cal V}$: $V = \sqrt{p_1 p_2 p_3}$. Dynamical equations can be obtained using Hamilton's equations:
\begin{equation}
\dot{p_i} \, = \, \{p_i,\mathcal{H}_{\rm cl}\} \, = \, -8 \pi G \gamma \frac{\partial \mathcal{H}_{\rm cl}}{\partial c_i} ~
\end{equation}
and
\begin{equation}
\dot{c_i} \, =\, \{c_i,\mathcal{H}_{\rm cl}\} \, = \, 8 \pi G \gamma \frac{\partial \mathcal{H}_{\rm cl}}{\partial p_i} ~
\end{equation}
where the `dot' represents the derivative with respect to the proper time $t$. Using the first equation one can compute  the directional Hubble rates, $H_i = \dot a_i/a_i$, such as
 \begin{equation*}
\label{hubbledef} H_1 \, = \,\frac{1}{2} \, \left(\frac{\dot{p_2}}{p_2} + \frac{\dot{p_3}}{p_3} - \frac{\dot{p_1}}{p_1}\right)
\end{equation*}
and show that
the connection component $c_i$ are related to $H_i$ in the classical theory as $ c_i \, = \, \gamma l_i H_i a_i$. These directional Hubble rates also define the expansion scalar $\theta$ and the shear scalar $\sigma^2$ for the comoving observers:
\be\label{eq:theta}
\theta \,  = \, \f{1}{V}\f{\d V}{\d t} \, =  \,  (H_1 + H_2 + H_3) \,
\ee
and
\be\label{sheardef}
\sigma^2  \, = \, \sum_{i=1}^3 \left(H_i - \theta \right)^2 \, = \, \f{1}{3}\left((H_1 - H_2)^2 + (H_2 - H_3)^2 + (H_3 - H_1)^2\right) ~.
\ee
In an isotropic spacetime Hubble rates in all directions are equal and the shear scalar vanishes.

Physical solutions of the Einstein's equations satisfy
the constraint $\mathcal{H}_{\rm cl}\approx 0$, which results in the following equation for the classical GR:
\be
H_1 H_2 + H_2 H_3 + H_3 H_1 \, = \, 8 \pi G \, \rho
\ee
where the energy density $\rho$ is computed as $\rho = \Hmatt/V$. Using the equations for $H_i$'s, it is possible to obtain the equation for the mean Hubble rate $H = \dot a/a$, which turns out to be of a similar form as the Friedmann equation in the flat isotropic model albeit with a presence of a terms proportional to the anisotropic shear:
\be\label{b1fried}
H^2 \, = \, \frac{8\pi G}{3} \, \rho \, + \, \frac{\Sigma^2}{a^6}
\ee
where $\Sigma^2 :=\f{1}{6}\sigma^2 a^6$. In the classical theory, $\Sigma^2$ turns out to be a constant of motion. Note that due to the presence of positive definite shear scalar, the energy density can take negative values without Hubble rate becoming imaginary. This is in contrast to the isotropic limit of the above equation, corresponding to the $k=0$ FRW model, where negative energy densities are not allowed when Hubble rate is real. Using above dynamical equations, one finds that at vanishing scale factors, $\rho$, $\theta$ and  $\sigma^2$ diverge. These lead to the divergence in curvature invariants and the break down of geodesic evolution at the singularities.\footnote{To give an example, we recall that the  square of the Weyl curvature $C_{\alpha \beta \mu \nu}$ can be expressed in terms of its electric ($E_{\alpha \beta}$) and magnetic ($H_{\alpha \beta}$) parts relative to a unit time-like vector field $({\bf u})$: $E_{\alpha \beta} = C_{\alpha \gamma \beta \delta} u^\beta u^\delta$, $H_{\alpha \beta} = {}^*C_{\alpha \gamma \beta \delta} u^\gamma u^\delta$, as  $C_{\alpha \beta \mu  \nu} C^{\alpha \beta \mu \nu} = 8 (E_{\alpha \beta} E^{\alpha \beta} - H_{\alpha \beta} H^{\alpha \beta})$ (see for eg. \cite{ellis-cosmo}). For the Bianchi-I model, $H_{\alpha \beta} = 0$ and  $E_{\alpha \beta} = {}^{(3)} R_{\alpha \beta} + \frac{\theta}{3} \sigma_{\alpha \beta} - (\sigma^\gamma_\alpha \sigma_{\gamma \beta} - \frac{2}{3} \sigma^2 \delta_{\alpha \beta})$ when anisotropic stress vanishes. Thus, a divergence in expansion and shear scalars, leads to a divergence in the square of the Weyl curvature.}

Let us now discuss the dynamics in LQC based on the effective Hamiltonian approach based on the embedding method \cite{jw,vt,psvt}.\footnote{Another approach is the truncation method \cite{trunc}, also based on geometric formulation of quantum mechanics. In contrast to the embedding approach which can be regarded as analogous to the variational methods, the truncation approach is on the lines of order by order perturbation theory. Though the method is more systematic, one needs to exercise a lot of care in dealing with truncation errors. Unlike the embedding approach, the modified equations from this method have not been widely tested for compatibility with the underlying quantum evolution.}  This approach is based on geometrical formulation of quantum mechanics \cite{aa_schilling}, where the space of quantum states is regarded as an infinite dimensional quantum phase space, $\Gamma_Q$, equipped with a symplectic structure provided by the Hermitian inner product on the Hilbert space. Using a judicious choice of states, one then finds an embedding of the finite dimensional classical phase space $\Gamma$. If this embedding is approximately preserved under the flow generated by the quantum Hamiltonian vector field, then up to the order of approximation the embedding is considered faithful, and one obtains effective equations which incorporate quantum corrections. In LQC, this procedure has been successfully carried out for a variety of models, including a dust dominated universe \cite{jw}, isotropic model with a massless scalar field \cite{vt} and isotropic model with arbitrary matter \cite{psvt}. The resulting effective Hamiltonian leads to modified Einstein's equations which have been extensively compared with the true quantum evolution using numerical simulations with states which are semi-classical at late times \cite{aps2,aps3,apsv,bp,kv,ap}. These analyses show that the modified Einstein's equations obtained from the effective Hamiltonian in LQC are in excellent agreement with the quantum evolution for all the models considered so far. Thus, suggesting that the effective Hamiltonian approach in LQC may have validity for a large class of spacetimes.

The effective Hamiltonian constraint for Bianchi-I spacetime in LQC for lapse $N=V$ is given as \cite{cv,csv,awe2}
\be
\Heff \, = \, -\frac{1}{8 \pi G \gamma^2}\left(\sinamu \sinbmu p_1 p_2 +\mbox{cyclic terms}\right) \, + \, \Hmatt V ~,
\ee
where $\bar \mu_i$ are given by
\be
\bar\mu_1=\lambda \sqrt{\f{p_1}{p_2 p_3}}; \quad \bar\mu_2=\lambda \sqrt{\f{p_2}{p_1 p_3}}; {\rm and} \quad \bar\mu_3=\lambda \sqrt{\f{p_3}{p_1 p_2}} ~
\ee
and $\lambda^2=4\sqrt{3}\pi \gamma \lp^2$. Note that $\Hmatt$ here corresponds to the matter Hamiltonian as obtained from the Fock quantization. The relationship of $\bar \mu_i$ with the triads is a consequence of the way area of the loops along which holonomies of connection are computed relates with the eigenvalue of the area operator in LQG.\footnote{It is important to note that at the current stage of research, the imposition of the minimum area of the loop as it appears in LQG is an external input in LQC. Since a derivation of LQC from LQG is yet to be performed, LQC is a quantization of cosmological models based on LQG, rather than the cosmological sector of LQG.}  The minimum allowed area of such a loop is labeled by $\lambda^2$. It is to be noted that above functional dependence of $\bar \mu_i$ on the triads is unique, in the sense that any other choice leads to resulting physics being affected by the rescaling of the lengths of the edges of the fiducial cell and also by change in its shape \cite{cs09}. Further, the only contribution from quantum geometry in the effective Hamiltonian constraint results from the modifications originating from the non-local field strength. The inverse volume modifications are absent in the above constraint, since we are considering a non-compact spatial manifold.

Using Hamilton's equations, we can compute the time variation of triads. In terms of proper time, we get
\be
\label{p1dot}\dot{p_1}=\frac{p_1}{\gamma \lambda} \left(\sinb+\sinc \right)\cosa
\ee
and similarly for $p_2$ and $p_3$, and the connection components. From these, we can obtain the equations for the directional Hubble rates, such as
 \be \label{H_bianchi1}
H_1  = \frac{1}{2\gamma \lambda}\left( \sin{(\mua c_1-\mub c_2)}+\sin{(\mub c_2+\muc c_3)}+\sin{(\mua c_1-\muc c_3)} \right)
\ee
Using these equations, one finds that that unlike the classical theory, $c_i \neq \gamma l_i H_i a_i$. Further, it is straightforward to show,  that $\Sigma^2$ is not a constant of motion in LQC \cite{cv,csv}.

In contrast to the classical theory, the directional Hubble rates are bounded in LQC, with a maximum value given by $H_{i, \mathrm{max}} = 3/2 \gamma \lambda$. The resulting expression for the expansion scalar yields
\be
\theta = \f{1}{2\gamma\lambda}\left( \sin{(\mua c_1+\mub c_2)}+\sin{(\mub c_2+\muc c_3)}+\sin{(\mua c_1+\muc c_3)} \right)
\ee
which has the following maxima
\be
\theta_{\mathrm{max}} = \f{3}{2 \gamma \lambda} \approx \f{2.78}{\lp} ~.
\ee
The boundedness of the Hubble rates points to the non-singular bounces in LQC. These bounces, unlike the isotropic case, do not occur at a fixed value of energy density. Due to the presence of anisotropies, bounce occurs at different values of $\rho$ and $\sigma^2$, which are determined by the initial conditions (see for example, Ref. \cite{csv}). In order to find the maximum values, we first note that the vanishing of the Hamiltonian constraint, $\Heff \approx 0$, gives
\be
\label{rhob1} \rho=\frac{1}{8 \pi G \gamma^2 \lambda^2}\left(\sina \sinb + \mbox{cyclic terms}\right) ~,
\ee
which implies that $\rho$ has a universal maxima, independent of any energy conditions, as:
\be
\rho_{\rm max}=\frac{3}{8 \pi G \gamma^2 \lambda^2}\approx 0.41 \rho_{\rm Pl} ~.
\ee
Thus, the maxima of energy density in Bianchi-I model turns out to be the same as in the isotropic models in LQC. Note that for an arbitrary choice of initial conditions in Bianchi-I spacetime, is never achieved at a bounce of the directional scale factor. This bound is only saturated when anisotropies vanish.

Similarly, using eq. (\ref{H_bianchi1}) (and equations for $H_2$ and $H_3$) in (\ref{sheardef}), it is straightforward to obtain the expression for shear scalar, which as the one for $\theta$ and $\rho$ turns out to be composed of bounded functions:
\ba
\label{shearb1} \sigma_{\rm I}^2&=& \nonumber \frac{1}{3 \gamma^2 \lambda^2 }\Bigg[ (\cosb(\sina+\sinc)- \cosa(\sinb+\sinc))^2  \\ && +~\mbox{~cyclic terms}~\Bigg] ~
\ea
(where the subscript $I$ is used to differentiate this expression from shear scalars in subsequent models considered here).
The shear scalar has a global maxima
\be
 {\sigma_{\rm I}^2}_{\rm max}=\frac{10.125}{3 \gamma^2 \lambda^2} \approx \f{11.57}{\lp^2}
\ee
at  $\bar \mu_1 c_1 = \pi/6, \bar \mu_2 c_2 =  \pi/2 ~{\rm and}~ \bar \mu_3 c_3 = 5\pi/6$. Interestingly, the maxima of shear scalar is reached when energy density is itself close to the Planckian value. At above values of $\bar \mu_i c_i$, the energy density turns out to be $\rho \approx 0.4167 \rho_{\rm crit}$. In the case of Bianchi-I vacuum spacetime ($\rho = 0$), depending on the initial conditions, the shear scalar can attain one of the two (local) maximas: $\sigma^2_{\mathrm{max}} = 2/\gamma^2 \lambda^2$ at  $\bar \mu_1 c_1 = \pi/2, \bar \mu_2 c_2 =  \pi ~{\rm and}~ \bar \mu_3 c_3 = 0$, or $\sigma^2_{\mathrm{max}} = 6.3/3\gamma^2 \lambda^2$ at  $\bar \mu_1 c_1 = -0.339837, \bar \mu_2 c_2 =  \pi/2 ~{\rm and}~ \bar \mu_3 c_3 = 5 \pi/6$.

The existence of upper bounds on energy density, expansion scalar and shear scalar for arbitrary matter indicate a resolution of various singularities in the loop quantization of Bianchi-I spacetime.\footnote{Strictly speaking, a stronger statement can be made for the  isotropic models  in LQC. This is due to the reason that in the isotropic case, availability of an exactly soluble model \cite{acs}, allows one to prove existence of bound on energy density for a dense set of states in the physical Hilbert space. Where as here,  bounds are derived  assuming an effective Hamiltonian which a priori assumes coherent states. Thus, it is possible that above bounds, and similarly those derived later for Bianchi-II and Bianchi-IX models, are not necessarily strict bounds for {\it all} the states in the physical Hilbert space.} This is a direct feature of the underlying quantum geometry captured via $\lambda^2$. In the limit, this parameter goes to zero, one recovers the classical divergence of $\rho$, $\theta$ and $\sigma^2$. this feature is missing in the classical theory. The upper bounds of these quantities are completely generic, obeyed by all types of matter. As we will discuss in the next sections, this generality is absent in Bianchi-II and Bianchi-IX models.

\section{Bianchi-II model}
Bianchi-II spacetimes are more general than the Bianchi-I spacetimes in the sense that they have a  non-vanishing intrinsic curvature.  Unlike the Bianchi-I model,  only two of the three Killing fields commute with each other on the spatial manifold. Further, the spacetime lacks an isotropic limit. In the following we consider the case of a non-compact Bianchi-II spacetime. Since the procedure of symmetry reduction, introduction of a cell ${\cal V}$ and details of the symplectic structure follow closely as in the Bianchi-I model, we will skip the discussion of this part. For a detailed discussion on these for Bianchi-II model, we refer the reader to Refs. \cite{awe3}.


In terms of Ashtekar variables, the classical Hamiltonian constraint for lapse $N=V$, can be written as
\be
\mathcal{H}_{\rm cl}=-\frac{1}{8 \pi G \gamma^2 }\left[p_1p_2c_1c_2 + \mbox{cyclic terms}\right]-\frac{1}{8 \pi G \gamma^2 }\Bigg[\alpha \, p_2p_3c_1 -(1+\gamma^2){\left(\frac{\alpha p_2p_3}{2p_1}\right)^2}\Bigg]+\Hmatt V
\ee
where $\alpha$ is related to the structure constants of the Lie algebra corresponding to the Killing fields.\footnote{The Killing fields $\mathring{\xi^a_i}$ satisfy $[\mathring{\xi_i}, \mathring{\xi_j}] = \mathring{C}_{ij}^k \mathring{\xi_k}$, with the only non-zero structure constant being $\mathring{C}^1_{23} = \tilde \alpha$. This defines $\alpha$ as $\alpha := (l_2 l_3/l_1) \tilde \alpha$, where $l_i$ refer to the edge lengths of the fiducial cell ${\cal V}$.} Using Hamilton's equations, it is straightforward to obtain the classical dynamical equations for the time variation of $p_i$ and $c_i$. As an example:
\be\label{p1_b2}
\f{\d p_1}{\d \tau} = \f{1}{\gamma} \left(p_1 (c_2 p_2 + c_3 p_3) + \alpha \, p_2 p_3\right) .
\ee
However, unlike the Bianchi-I model, it is not possible to use these equations to obtain an analog of the generalized Friedmann equation (\ref{b1fried}) for the mean Hubble rate. A detailed analysis of these dynamical equations reveal a singularity when the triads vanish.

We now consider the effective Hamiltonian for the Bianchi-II model \cite{awe3}. For the same choice of the lapse function as above, it is given as
\ba
\label{heffb2} \Heff & = & \nonumber - \, \frac{p_1p_2p_3}{8 \pi G \gamma^2 \lambda^2}\left[\sina \sinb +\mbox{cyclic terms}\right] \\ && -\, \frac{1}{8 \pi G \gamma^2 }\left[\frac{\alpha}{\lambda} \frac{(p_2 p_3)^{3/2}}{\sqrt{p_1}} \sina-\frac{\alpha^2(1+\gamma^2)}{4}\left({\frac{p_2 p_3}{p_1}}\right)^2\right] ~
+\Hmatt V~
\ea
which is a sum of the effective Hamiltonian in Bianchi-I model  and the terms originating from the presence of the spatial curvature. Imposing the
the constraint $\Heff\approx 0$ we obtain
\be
\label{rhob2} \rho = \frac{1}{8 \pi G \gamma^2 \lambda^2}\left[\sina \sinb +\mbox{cyclic terms}\right]+\frac{1}{8 \pi G \gamma^2 }\left[\frac{x}{\lambda}\sina-\frac{(1+\gamma^2)x^2}{4}\right] ~
\ee
where as in the Bianchi-I model, $\rho = \Hmatt/V$ and $x$ is defined as
\be
x = \alpha \sqrt{\f{p_2 p_3}{p_1^3}} ~.
\ee
The right hand side of the eq. (\ref{rhob2}) has a global maxima at $x=\frac{2}{(1+\gamma^2)\lambda}$ and $\sin(\bar{\mu}_ic_i)=1$. This results in a  maximum allowed value of the energy density as \cite{awe2}:
\be
\label{rhoboundb2} \rho \leq \rmax=\frac{3+{(1+\gamma^2)}^{-1}}{8 \pi G \gamma^2 \lambda^2} \approx 0.54 \rho_{\rm Pl} ~.
\ee
The above upper bound on $\rho$ is the global maxima.  
It is straightforward to see that eq. (\ref{rhob2}) can give rise to large negative energy densities depending on the values of triads.\footnote{As noted below eq.(\ref{b1fried}), in the anisotropic spacetimes, negative values of energy density do not necessarily cause problems with the reality of the expansion scalar.}  This behavior is depicted in Fig. 1. A possible way to obtain a lower bound on the energy density is by imposing energy conditions on the matter content. As an example, if we assume weak energy condition (WEC), then $\rho \geq 0$. This procedure is similar to imposing energy conditions in GR to eliminate solutions with negative energy density. However, unlike GR, where even after imposing energy conditions, energy density is not bounded above, in LQC imposition of an energy condition on $\rho$ does not affect the global maxima of $\rho$ which as we have shown above arises because of the underlying quantum geometric effects. 
\begin{figure}[tbh!]
\includegraphics[angle=0,width=0.5\textwidth]{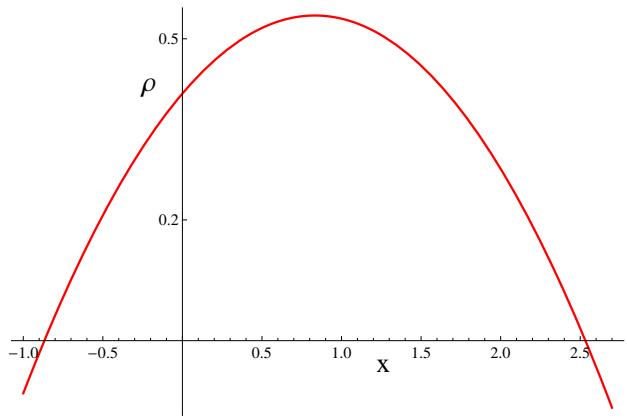}
\caption{The plot (in Planck units) shows the variation of energy density as a function of $x$ in the loop quantization of non-compact Bianchi-II model when $\sin(\bar{\mu}_ic_i)=1$ is substituted in eq. (\ref{rhob2}). }
\end{figure}

Using (\ref{heffb2}), one can derive the modified dynamical equations using Hamilton's equations. For the triad components we obtain
\begin{equation*}
\f{\d p_1}{\d \tau}=\f{1}{\gamma}\left(\frac{p_1^2}{\bar\mu_1}\left(\sinb +\sinc \right)+\alpha p_2 p_3\right)\cosa
\end{equation*}
which yields the classical equation (\ref{p1_b2}) in the limit $\lambda^2 \rightarrow 0$. In terms of the proper time, this equation can be written as
\be
\frac{\dot{p_1}}{p_1}=\frac{1}{\gamma \lambda}\left( \sinb +\sinc + \lambda x \right)\cosa ~.
\ee
Equations for the time variation of other triads can be derived in a similar way, and they turn out to be:
\begin{gather}
\frac{\dot{p_2}}{p_2}=\frac{1}{\gamma \lambda}\left( \sina +\sinc \right)\cosb \\
\frac{\dot{p_3}}{p_3}=\frac{1}{\gamma \lambda}\left( \sina +\sinb \right)\cosc ~.
\end{gather}

Using these equation we can obtain the directional Hubble rates, $H_i$, and the expansion scalar $\theta$ using eq.(\ref{eq:theta}) which becomes
\be\label{theta_b2}
\theta=\f{1}{2\gamma\lambda}\left(\sin(\mua c_1+\mub c_2)+\sin(\mub c_2+\muc c_3)+\sin(\muc c_3+\mua c_1)+\lambda x \cosa\right) ~.
\ee
Unlike the expansion scalar in the Bianchi-I model,  $\theta$ in Bianchi-II model is generically unbounded because of the divergence in $x$  as $p_1 \rightarrow 0$ or ($p_2, p_3 \rightarrow \infty$). However, if one imposes energy conditions demanding that the energy density be bounded below, then $x$ can not grow beyond a maximum value, and $\theta$ turns out to be bounded. For matter satisfying WEC, the maximum allowed value of $\theta$ is given by,
\be
\theta_{\rm max} \approx \f{6.05}{2\gamma\lambda} \approx \f{5.60}{l_{\rm Pl}}~.
\ee
This upper bound can be obtained by finding the maxima of eq.(\ref{theta_b2}) numerically by optimizing various variables and occurs
at $\mua c_1=0.642$, $\mub c_2=0.982$, $\muc c_3=0.982$ and $x=1.717$.

\vskip0.5cm

Finally, the expression for the shear scalar in Bianchi-II model, $\sigma^2_{\rm II}$ can be obtained using (\ref{sheardef}), which after a straightforward calculation yields,
\ba
\label{shearb2} \sigma_{\rm II}^2 &=& \nonumber \sigma_{\rm I}^2+\frac{1}{3 \gamma^2 \lambda^2} \Bigg[2 \lambda^2 x^2 \cos^2(\bar \mu_1 c_1) + 2\lambda x \cosa (2(\sinb+\sinc)\cosa  \\ && -(\sina+\sinc)\cosb  -(\sinb+\sina)\cosc) \Bigg]~
\ea
where $\sigma_{\rm I}^2$ is given by eq.(\ref{shearb1}). This term is bounded, as shown in Sec. II. However, the term with square parenthesis is an increasing quadratic function in $x$. Thus,  $\sigma^2_{\rm II}$ does not have an upper bound.  The shear scalar for non-compact Bianchi-II model is devoid of a generic maxima in contrast to the Bianchi-I spacetime. In order to obtain a bound on $\sigma^2_{\rm II}$, one needs to assume a lower bound for the energy density, as for the expansion scalar.  Imposing WEC, the maximum value of $\sigma^2_{\rm II}$ turns out to be
\be
{\sigma_{\rm II}^2}_ {\mathrm{~max}}\approx \frac{57.58}{3\gamma^2\lambda^2}\approx\f{65.82}{l_{\rm Pl}^2} ~.
\ee
Note that the maximum of the shear scalar, as that of the energy density and the expansion scalar in the non-compact Bianchi-II model turns out to be different from the Bianchi-I model.\\

It is to be emphasized that the bound on the shear scalar is sensitive to the energy conditions, i.e. if one imposes a different energy condition, the allowed range for the parameter $x$ would be different and so would be the upper bound on the shear scalar. WEC is followed by almost all types of the matter and it provides an adequate estimation of the bounds on shear and energy density. Thus, WEC has two roles: first to give a lower bound on $\rho$ and the second  to provide the corresponding upper bounds on $\theta$  and $\sigma^2_{\rm II}$. Further, the upper bounds on $\theta$ and $\sigma^2$ are saturated in the dynamical evolution, and the bounce of directional scale factors  occur before these values are reached. As in the case of Bianchi-I model, this is tied to the interplay of the Ricci and the Weyl parts  of the spacetime curvature in the Bianchi-II model.

\section{Bianchi-IX Model}
The spatial manifold for the Bianchi-IX model has a compact topology $S^3$ whose radius with respect to the fiducial metric is chosen as $a_r = 2$. The volume of fiducial cell is given by
$V_o =: \ell_o^3 = 2 \pi^2 a_r^3 = 16 \pi^2$. In contrast to the Bianchi-I and Bianchi-II spacetimes, none of the Killing vectors commute with each other in the Bianchi-IX model. In this spacetime, the interplay of intrinsic curvature and anisotropies is also much richer in comparison to the Bianchi-II model. Classical dynamics can exhibit a Mixmaster behavior as singularities are approached. In the isotropic limit, one recovers the classical dynamics of the $k=1$ FRW model.

Let us first consider the classical Hamiltonian constraint for the Bianchi-IX model. With lapse $N = V$, in terms of the connections and triads, it is given by
\ba \label{b9hcl}
\mathcal{H}_{\rm cl}&=&-\f{1}{8 \pi G\gamma^2}\Bigg(p_1p_2c_1c_2+p_2p_3c_2c_3+p_3p_1c_3c_1+\f{2  \ell_o}{2}\left(p_1p_2c_3+p_2p_3c_1+p_3p_1c_2\right)\nonumber \\ &&+ \f{\ell_o^2}{4}\left(1+\gamma^2\right)\left[2p_1^2+2p_2^2+2p_3^2-\left(\f{p_1p_2}{p_3}\right)^2-\left(\f{p_2p_3}{p_1}\right)^2-\left(\f{p_3p_1}{p_2}\right)^2\right]\Bigg)+\Hmatt V~ \nonumber \\ &&
\ea
which using Hamilton's equation for motion, leads to
\be
\f{\d p_1}{\d \tau}=\f{p_1}{\gamma}\Bigg[p_2c_2+p_3c_3+\ell_0 \f{p_2p_3}{p_1}\Bigg]
\ee
and similarly for $p_2$ and $p_3$, and the connection components. In this case, one can use these equations to derive a generalized Friedmann equation, as in the Bianchi-I model, and it turns out to be:
\be
H^2=\f{8\pi G}{3}\rho+\f{1}{6}\sigma^2
-\f{\ell_o^2}{12p_1p_2p_3}\Bigg[2\left(p_1^2+p_2^2+p_3^2\right)-\left(\f{p_1p_2}{p_3}\right)^2-\left(\f{p_2p_3}{p_1}\right)^2-\left(\f{p_3p_1}{p_2}\right)^2\Bigg] ~.
\ee
The Hubble rate diverges as the singularities are approached in the Bianchi-IX spacetime. Further, due to the presence of intrinsic curvature, singularities can occur both in past and the future evolution, as in the $k=1$ universe. As shown in Appendix A,
in the isotropic limit, one recovers the dynamical equations for the $k=1$ model.

As in the case of the loop quantization of the Bianchi-II model, it is not possible to express the field strength operator in terms of holonomies along closed loops. To overcome this difficulty, a quantization has been proposed, along the lines of Bianchi-II model, following the connection operator approach \cite{we1}. In the following we analyze the physics resulting from the effective Hamiltonian constraint of this quantization. Since the underlying manifold is spatially compact, the resulting effective Hamiltonian constraint contains modifications originating from both the non-local nature of the field strength operator and the eigenvalues of the inverse triads.  Let us first analyze some features of the resulting physics for the effective Hamiltonian constraint of Bianchi-IX model, if inverse triad modifications are ignored (as in the analysis of Ref. \cite{we1}). We denote this effective Hamiltonian constraint by $\tilde {\cal H}_{\mathrm{eff}} $ in order to distinguish it from the Hamiltonian constraint in eq. (\ref{effhamb9}) where the inverse triad corrections are included. For the lapse $N=V$, the effective Hamiltonian constraint is given by \cite{we1}:
\begin{multline}\label{b9hlqc1}
\tilde {\cal H}_{\mathrm{eff}} =\frac{p_1 p_2 p_3}{8 \pi G \gamma^2\lambda^2}\left[\sina \sinb + \mbox{cyclic terms} \right] -\frac{\ell_o}{8 \pi G \gamma^2 \lambda}\left[\frac{(p_1 p_2)^{3/2}}{\sqrt{p_3}} \sinc +\mbox{cyclic terms}\right] \\
-\frac{{\ell_o}^2}{32 \pi G \gamma^2}{(1+\gamma^2)}\left[2(p_1^2+p_2^2+p_3^2)-\left(\left(\frac{p_2 p_3}{p_3}\right)^2+\mbox{cyclic terms}\right)\right] +\Hmatt  V~.
\end{multline}
The vanishing of this constraint leads to the following expression of energy density
\ba
\label{rhob9} \rho&=&\frac{1}{8 \pi G \gamma^2 \lambda^2}\left[\sina \sinb + \mbox{cyclic terms} \right] +\frac{\ell_o}{8 \pi G \gamma^2 \lambda}\left[{\xc} \sinc +\mbox{cyclic terms}\right] \nonumber \\ &&
+\frac{{\ell_o}^2(1+\gamma^2)}{32 \pi G \gamma^2}\left[\left(2\xbxc-\frac{p_2 p_3}{p^3_1}\right)+\mbox{cyclic terms}\right] ~.
\ea
Utilizing the boundedness properties of the trigonometric functions, this  expression  implies that
\be
\label{rhomax} \rho \leq \frac{3}{8 \pi G \gamma^2 \lambda^2}+\frac{\ell_o}{8 \pi G \gamma^2 \lambda}\left[x_1+x_2+x_3\right] +\frac{\ell_o^2 (1+\gamma^2)}{32 \pi G \gamma^2}\left[2(x_1 x_2+x_2 x_3+x_3 x_1)-{(x_1^2+x_2^2+x_3^2)}\right]
\ee
where $x_1=\xa, x_2=\xb, \mbox{and } x_3=\xc$.

The first term in eq. (\ref{rhomax}) is the maximum energy density obtained in Bianchi-I spacetime. It is the behavior of the second and the third term, which determines the boundedness of energy density from the above effective Hamiltonian constraint. In order that the energy density to have a maxima,  $\rho(x_1,x_2,x_3)$ 
should have a  viable simultaneous solution to the following system of equations:
\be
\frac{\partial \rho{(x_1,x_2,x_3)}}{\partial x_1}=0; \quad \frac{\partial \rho{(x_1,x_2,x_3)}}{\partial x_2}=0; \quad \frac{\partial \rho{(x_1,x_2,x_3)}}{\partial x_3}=0 ~.
\ee

Solving these equations, one obtains the following condition:
\be
\label{cond_rho} \frac{6}{\ell_o \lambda (1+\gamma^2)}+x_1+x_2+x_3=0 ~.
\ee
Thus a physical solution is allowed only when at least one of $x_i$ is \emph{negative}. However, by definition,  all of $x_i$ are \emph{positive}.\footnote{The unboundedness argument given here is valid irrespective of the choice of orientation of triads, which has been here fixed to be positive.} The energy density resulting from $\tilde {\cal H}_{\mathrm{eff}}$ does not have a  maxima, unlike the Bianchi-I and Bianchi-II models in the absence of inverse triad modifications. On analyzing the modified dynamical equations resulting from $\tilde {\cal H}_{\mathrm{eff}}$, we find that the expansion and the shear scalars are also unbounded. \\

However, as discussed above the effective Hamiltonian constraint $\tilde {\cal H}_{\mathrm{eff}}$ is incomplete as it lacks the contribution from inverse triad corrections in the Bianchi-IX model which is spatially compact.  To conclude whether the energy density in Bianchi-IX model has an upper bound, it is necessary to include these modifications. To introduce these modifications in the corresponding effective Hamiltonian constraint in LQC for such terms,  we consider eigenvalues $f(p_i)$ of the operator $\widehat{{p_i}^{-1/2}}$ and substitute them in place of inverse triad terms.\footnote{The strategy is same as in the isotropic models and is based on the identities first proved by Thiemann in LQG \cite{tt}. For an up to date discussion of these modifications in isotropic models and related subtleties for non-compact models, see Ref. \cite{as}.} As an example, the eigenvalues for the inverse triad operator $\widehat{{p_1}^{-1/4}}$ turn out to be \cite{awe2}

\begin{equation}
\widehat{{p}_1^{-1/4}} |p_1,p_2,p_3\rangle= f(p_1) |p_1, p_2, p_3 \rangle
\ee
where $|p_1,p_2,p_3\rangle$ denote eigenstates of the volume operator, and $v=\frac{2}{4 \pi \gamma \lambda \lp^2} \sqrt{p_1 p_2 p_3}$ and
\be
f(p_1)=\frac{2}{4 \pi \gamma \lambda \lp^2} (p_2 p_3)^{1/2}\Bigg[\sqrt{|v+1|}- \sqrt{|v-1|}\Bigg]^2 ~.
\ee
Similarly, one can derive expressions for other inverse triad operators.
With these modifications, the effective Hamiltonian constraint becomes:
\begin{eqnarray}
\Heff \label{effhamb9} \, &=& \nonumber \, - \, \frac{p_1p_2p_3}{8 \pi G \gamma^2 \lambda^2}\Bigg[\sina \sinb+ \mathrm{cyclic} ~~ \mathrm{terms}\Bigg]  \\ && \, \nonumber -\frac{\varepsilon \ell_o}{8 \pi G \gamma^2 \lambda}\Bigg[{(p_1p_2)}^{3/2} f(p_3)\sinc +{(p_2p_3)}^{3/2} f(p_1)\sina+{(p_3p_1)}^{3/2} f(p_2)\sinb\Bigg]  \\ && \nonumber -\frac{\ell_o^2 (1+\gamma^2)}{32 \pi G \gamma^2}\Bigg[2(p_1^2+p_2^2+p_3^2)-{(p_1p_2)}^2 f(p_3)^4-(p_2p_3)^2 f(p_1)^4-(p_3p_1)^2f(p_2)^4\Bigg] +\Hmatt V~\\&&
\ea
where the inverse triad modifications also contribute to $\Hmatt$ if one considers matter Hamiltonian containing inverse powers of the scale factor.

We now analyze the behavior of  energy density. Due to the inverse triad modifications, an ambiguity in its definition arises. Using the effective Hamiltonian constraint, one can define the energy density $\rho$ as $\rho = \Hmatt/V$, where the $\Hmatt$ includes modifications due to inverse triad operators.  However, one can also define energy density such that it agrees with the eigenvalues of $ \widehat{V^{-1} {\cal H}_{\mathrm{matt}}}$ (suitably symmetrized). We label this energy density as $\rho_q$. In the absence of inverse volume modifications, $\rho$ and $\rho_q$ are equal to each other. However, when these are present, $\rho_q$ and $\rho$ can behave in a qualitatively different way, at small volumes.\footnote{In LQC, physics has been analyzed using both of these definitions. For a comparison of some of the features, see Ref. \cite{ps05}.} As it turns out, $\rho$ and $\rho_q$, indeed have qualitative differences in the loop quantization of Bianchi-IX model.

Using the effective Hamiltonian, a division by volume, leads to the following inequality for the energy density $\rho$ for the physical solutions:
\ba \label{densityb9_1}
\rho &\leq& \frac{3}{8 \pi G \gamma^2 \lambda^2}+\frac{\ell_o}{8 \pi G \gamma^2 \lambda}\Bigg[\frac{\sqrt{p_1p_2}}{p_3} f(p_3)+\frac{\sqrt{p_2p_3}}{p_1} f(p_1)+\frac{\sqrt{p_3p_1}}{p_2} f(p_2)\Bigg]  \\ && \nonumber +\frac{\ell_o^2 (1+\gamma^2)}{32 \pi G \gamma^2}\Bigg[2\left(\frac{p_1}{p_2p_3}+\frac{p_2}{p_1p_3}+\frac{p_3}{p_2p_1}\right)-\frac{p_1p_2}{p_3} f(p_3)^4-\frac{p_2p_3}{p_1} f(p_1)^4-\frac{p_3p_1}{p_2} f(p_2)^4\Bigg]~.
\ea

\begin{figure}[tbh!]
\begin{center}
\hspace{0.8cm}
\includegraphics[angle=0,width=0.5\textwidth]{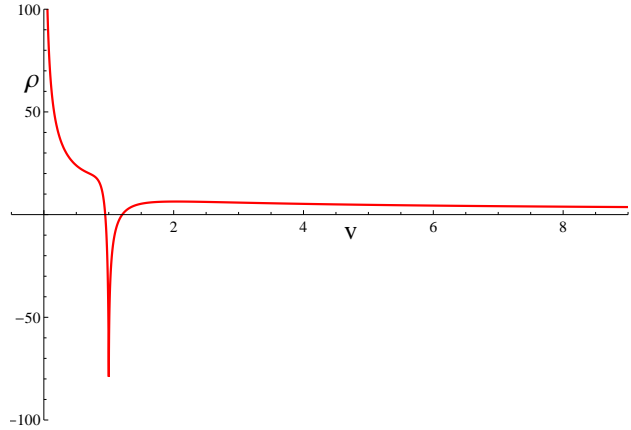}
\caption{Behavior of energy density $\rho$ as a function of $v$ is shown (in Planck units).}
\end{center}
\end{figure}

The behavior of energy density with respect to $v$ is shown in Fig. 2. We find that the energy density $\rho$ does not have a global maxima if the effective Hamiltonian description is assumed to be valid for the entire range of $v$. However, in the case when all the triads approach the classical singularity at the same time  i.e. the singularity is isotropic, then for the range $v > 1$ there exists a local maxima at $p_i \approx 3.634 \lp^2$ , given by
\be
\rho_{\mathrm{max}} \approx 6.34 \rho_{\mathrm{Pl}} ~.
\ee
Thus, for the energy density defined as the ratio of the matter Hamiltonian to the physical volume, inverse triad modifications do not suffice to control the divergence as $v \rightarrow 0$.

Let us now analyze the behavior of $\rho_q$. In this case, the effective Hamiltonian constraint yields
\ba
\rho_{\rm q} &\leq& g(v)\Bigg[ \frac{3\sqrt{p_1p_2p_3}}{8 \pi G \gamma^2 \lambda^2}+\frac{\ell_o}{8 \pi G \gamma^2 \lambda}\Bigg[\frac{p_1p_2}{\sqrt{p_3}} f(p_3)+\frac{p_2p_3}{\sqrt{p_1}} f(p_1)+\frac{p_3p_1}{\sqrt{p_2}} f(p_2)\Bigg]  \\  && \, \nonumber+\frac{\ell_o^2 (1+\gamma^2)}{32 \pi G \gamma^2}\Bigg[2\left(\frac{p_1^{3/2}}{\sqrt{p_2p_3}}+\frac{p_2^{3/2}}{\sqrt{p_1p_3}}+\frac{p_3^{3/2}}{\sqrt{p_2p_1}}\right)-\frac{(p_1p_2)^{3/2}}{\sqrt{p_3}} f(p_3)^4 \\ && \, \nonumber-\frac{(p_2p_3)^{3/2}}{\sqrt{p_1}} f(p_1)^4-\frac{(p_3p_1)^{3/2}}{\sqrt{p_2}} f(p_2)^4\Bigg]\Bigg]~,
\ea
where $g(v)$ denotes the eigenvalue of the volume operator$\widehat{(1/V)}$
\be
g(v)=\f{1}{(2\pi\gamma\lambda \lp^2)^3}\left(p_1p_2p_3\right)\left[\sqrt{v+1}-\sqrt{v-1}\right]^6 ~.
\ee
 We find that there exists a global maxima at $p_i \approx2.109 \lp^2$, when the approach to classical singularity is isotropic, given by
\be
\rho_{\rm{q \,max}}\approx 11.74 \rho_{\rm Pl} ~.
\ee
The variation of $\rho_q$ with respect to $v$ is shown in Fig. 3. Note that this maximum value is higher than the one in Bianchi-I or Bianchi-II models. When the approach to the classical singularities is not isotropic, such as for pancake or cigar singularities, above bounds can be violated. This is evident from Fig. 4, where we see that {\it if} in the effective dynamics two of the triads vanish and the third diverges, {\it then} the energy density $\rho_q$ diverges. In all other cases, it is finite.

\begin{figure}[tbh!]
\includegraphics[angle=0,width=0.5\textwidth]{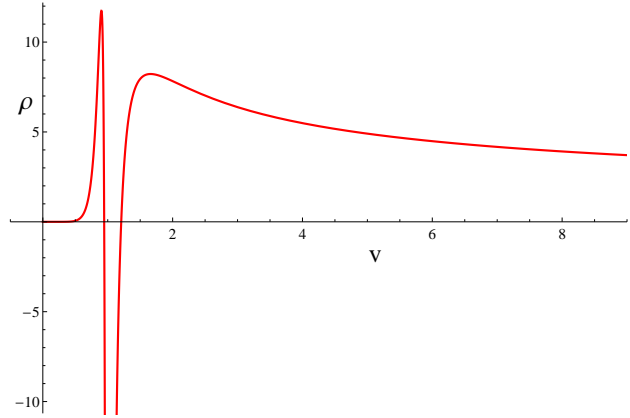}
\caption{Variation of energy density $\rho_{q}$ is shown versus $v$ in Planck units when the approach to classical singularities is isotropic. The minimum occurs at $-753.09 \rho_{\rm Pl}$. }
\end{figure}

Let us now obtain the dynamical equations for the triads and the expression for the expansion scalar. Using the effective Hamiltonian constraint (\ref{effhamb9}), one obtains using Hamilton's equations:
\be
\frac{\dot p_1}{p_1}=\frac{1}{\gamma \lambda}\Bigg[\sinb+\sinc+ \varepsilon \, \lambda \ell_o\frac{\sqrt{p_2p_3}}{p_1} f(p_1)\Bigg]\cosa
\ee
and similarly for $\dot p_2$ and $\dot p_3$. Using these, the expansion scalar can be computed as
\be
\theta=\f{1}{2\gamma \lambda}\Bigg[\left(\sin{(\mub c_2+\muc c_3)}+ \varepsilon \, \lambda \ell_o\cosa \frac{\sqrt{p_2p_3}}{p_1} f(p_1)\right) + \mbox{cyclic terms}\Bigg] ~.
\ee
The expansion scalar turns out to have a global maxima at $p_i \approx 2.258 \lp^2$, for the isotropic approach to classical singularity, given by
\be
\theta_{\rm max}\approx\f{47.72}{2\gamma\lambda}\approx \f{44.18}{\lp} ~.
\ee
whereas for more general approaches to singularity such as cigar or pancake like ones, there exists no such global maxima for $\theta$.

Note that unlike the Bianchi-II model, this is a global bound which does not depend on the imposition of any energy condition. Nevertheless, if one imposes WEC as in the Bianchi-II case, one get a lower value. The maximum allowed value of $\theta$ for matter satisfying WEC (in terms of $\rho_q$) is $\theta_{\mathrm{max}} = 30.84/2\gamma\lambda$.\\

\begin{figure}[tbh!]
\label{b9rho3d}
\includegraphics[angle=0,width=0.5\textwidth]{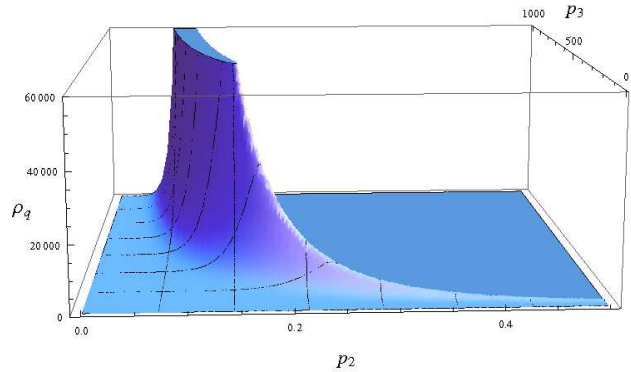}
\caption{Variation of energy density $\rho_{\rm q}$, in Planck units, is shown versus $p_2$  and $p_3$ (with $p_1=p_2$). The energy density diverges as two of the triads tend to zero while the other approaches infinity. For other trajectories $\rho_{\rm q}$ is finite.}
\end{figure}

We now consider the behavior of shear scalar in the Bianchi-IX model. Using Hamilton's equations, we obtain
\begin{multline}
\sigma_{\rm IX}^2=\f{1}{3\gamma^2\lambda^2}\Bigg[\Bigg( \left(\sinb+\sinc+\varepsilon \, \lambda \ell_o\frac{\sqrt{p_2p_3}}{p_1} f(p_1)\right)\cosa- \\  \left(\sina+\sinc+\lambda \ell_0\frac{\sqrt{p_1p_3}}{p_2} f(p_2)\right)\cosb \Bigg)^2+{\rm cyclic \quad terms}\Bigg] ~.
\end{multline}
For the isotropic approach to classical singularity, the shear scalar has a global maxima at $p_i \approx 2.258 \lp^2$, given by
\be
{\sigma_{\rm IX}^2}_{\rm max} \approx \f{2165.91}{3\gamma^2\lambda^2}\approx\f{2476.04}{\lp^2} ~.
\ee
For matter satisfying the WEC (for $\rho_q$), the maximum allowed value is given by $\smax \approx 690.98/(3\gamma^2\lambda^2)$. If the approach to the classical singularity is not isotropic, then the shear scalar can take values greater than the above value. This behavior can be seen from Fig. 5. There exist two cases where shear scalar can diverge in this model. {\it If} effective dynamics allows solutions where one triad vanishes and two diverge or where two of the triads vanish and the third one diverges, {\it then} the shear scalar can diverge. It is important to note that the existence of above divergent cases is based on the assumption that the effective Hamiltonian (\ref{effhamb9}) remains valid as above anisotropic singularities are approached. Since the validity of (\ref{effhamb9}) has not been tested with the underlying quantum evolution, it is possible that these cases originate in the regime where the effective dynamics resulting from (\ref{effhamb9}) may break down.

\begin{figure}[tbh!]
\label{shear3d}
\includegraphics[angle=0,width=0.5\textwidth]{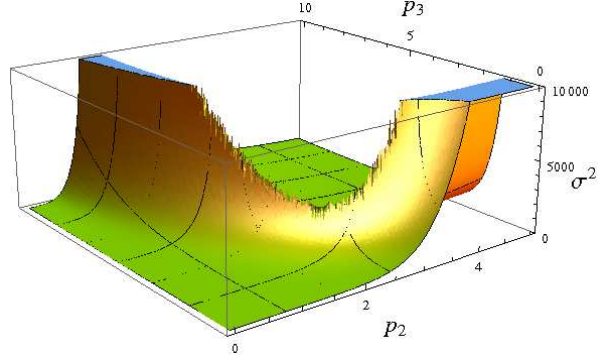}
\caption{Variation of shear scalar $\sigma^2$ for Bianchi-IX is shown versus $p_2$ and $p_3$ (with $p_1=p_2$). For isotropic approach to singularity there exists a local maxima.}
\end{figure}

Thus, we find that in the Bianchi-IX model, energy density $(\rho_q)$, expansion scalar and the shear scalar are all bounded for isotropic approach to singularity, if we include inverse triad modifications to the effective Hamiltonian constraint. However, energy density $(\rho)$ defined by taking a ratio of the matter Hamiltonian to the physical volume does
not have a global maxima. It is to be emphasized that  in the dynamical evolution these bounds are not saturated and in certain situations, bounce of scale factors can occur close to the values in Bianchi-I and Bianchi-II models.   \\

We conclude this section, with a discussion of the isotropic limit of the Bianchi-IX model. As mentioned earlier, in the classical theory, the limit is given by the
classical Hamiltonian constraint corresponding to the $k=1$ model. Considering the isotropic limit of classical Hamiltonian constraint (\ref{b9hcl}), one obtains
\be
\label{b9iso} \mathcal{H}_{\rm cl}^{\rm (iso)}=-\f{3 {p^{2}}}{8 \pi G \gamma^2}\left[\left(c+\f{\ell_o}{2}\right)^2+\f{\ell_o^2}{4}\gamma^2\right]+\mathcal{H}_{\rm matt} V ~.
\ee
This corresponds to the classical Hamiltonian constraint of $k=1$ model (eq. (\ref{k1ham})) with $\varepsilon = -1$. The pertinent question is whether the effective Hamiltonian constraint in Bianchi-IX model leads to the effective Hamiltonian constraint of the $k=1$ model in LQC. The quantization of $k=1$ model, based on expressing field strength in terms of holonomies over closed loops was performed in Ref. \cite{apsv}. A detailed analysis of the physics of this model reveals features similar to the $k=0$ model in LQC, with a bounce occurring at $\rho = \rho_{\mathrm{max}} \approx 0.41 \rho_{\mathrm{Planck}}$, and resolution of various strong curvature singularities \cite{sv}. In Appendix A, we discuss the effective Hamiltonian constraint corresponding to this quantization (eq. (\ref{k1hapsv})), and show that the expansion scalar in this quantization is  bounded above by a universal value.

Let us first consider the effective Hamiltonian constraint without the inverse triad modifications. The isotropic limit of ${\tilde{\cal H}}_{\mathrm{eff}}$ (eq.(\ref{b9hlqc1})), obtained by imposing $p_i = p$ and $c_i = c$, leads to 
\be
{\mathcal{\tilde H}^{(\rm iso)}_{\mathrm{eff}}}=-\f{3{p^{2}}}{8 \pi G \gamma^2}\Bigg[\f{\sin^2(\mb c)}{\bar{\mu}^2} + \ell_o  \f{\sin(\mb c)}{\bar{\mu}} +\f{\ell^2_o(1+\gamma^2)}{4}\Bigg]+\Hmatt V~.
\ee
This equation can be written as
\be
\label{b9isoquant} {\mathcal{\tilde H}^{(\rm iso)}_{\mathrm{eff}}}=-\f{3 p^{2}}{8 \pi G \gamma^2}\Bigg[\left(\frac{\sin(\bar{\mu}c)}{\bar{\mu}}+\f{\ell_o}{2}\right)^2 +\f{\ell_o^2\gamma^2}{4}\Bigg]+\Hmatt V~
\ee
which does not agree with (\ref{k1hapsv}).\footnote{It is straightforward to check that this is true irrespective of the choice of orientation of the triads.} Hence, the effective Hamiltonian constraint $\tilde {\cal H}_{\rm eff}$ in the isotropic limit does not yield the effective Hamiltonian constraint of the $k=1$ model in the quantization based on holonomies over closed loops \cite{apsv}.

It turns out that the isotropic limit of (\ref{b9hlqc1}) corresponds to the effective Hamiltonian constraint of an alternate quantization of $k=1$ model based on the connection operator approach (eq.(\ref{heff_k1conn})) discussed in Appendix A. This is straightforward to see using
(\ref{b9iso}) which under: $c \rightarrow \sin(\bar \mu c)/\bar \mu$, yields (\ref{b9isoquant}). However, as discussed in Appendix A, this quantization has the following drawback in comparison to the one in Ref. \cite{apsv}. If one assumes, the validity of the effective Hamiltonian constraint for all $v$, the expansion scalar turns out to be unbounded even after the inclusion of inverse triad modifications.\\

Finally, we consider the isotropic limit of the constraint (\ref{effhamb9}). Imposing $p_i = p$ and $c_i = c$, where $p$ and $c$ refer to the isotropic triad and connection variables, we obtain
\ba
{\cal H}_{\mathrm{eff}}^{\rm (iso)} &=& -\f{3p^3}{8\pi G\gamma^2\lambda^2}\left[\sin^2(\bar{\mu} c)\right]-\f{3\ell_0}{8\pi G\gamma^2\lambda}\left[p^3f(p)\sin(\bar{\mu c})\right] \\ && \, \nonumber-\f{3\ell_0^2(1+\gamma^2)}{32\pi G\gamma^2}\left[2p^2-p^2f(p)^4\right]+\Hmatt V
\ea
which agrees neither with the effective Hamiltonian constraint of Ref. \cite{apsv} nor the alternate quantization in the presence of inverse triad modifications. In summary, the effective Hamiltonian constraint of Bianchi-IX model in the isotropic limit does not yield the effective Hamiltonian constraint of the loop quantization of $k=1$ model performed using holonomies over closed loops \cite{apsv}. Its isotropic limit, ignoring inverse triad corrections, agrees with the alternate quantization of $k=1$ model \cite{ck}, in which the expansion scalar does not have a maxima.

\section{Discussion}

In this work, we investigate the behavior of energy density, the expansion and the shear scalars in Bianchi-II and Bianchi-IX models using the effective Hamiltonian approach in LQC.  These spacetimes
have a non-vanishing spatial curvature which leads to technical difficulties to carry out the loop quantization as in the Bianchi-I model. In particular, it is not possible to express field strength in terms of holonomies around closed loops such that the holonomies are almost periodic functions of connection components. The resulting holonomy operators are not well defined on the kinematical Hilbert space. To overcome this problem, a strategy was proposed in Ref. \cite{awe3}, which expressed field strength directly in terms of a non-local connection operator, defined via holonomies computed along open segments. In the connection operator approach, coefficients of the expression of field strength are fixed by demanding consistency with the Bianchi-I model \cite{awe3}. The resulting quantum Hamiltonian constraint turns out to be non-singular as in the Bianchi-I model. However, physics of these spacetimes was largely unexplored. First, it was not known whether the energy density and the expansion and shear scalars are bounded in these models. It was also not known whether they share same universal bounds as the Bianchi-I model and if one required extra conditions to obtain the boundedness of these physical quantities. Further, the important role played by inverse triad modifications in Bianchi-IX spacetime was so far unrecognized. It is important to emphasize that the bounds on $\rho$, $\theta$ and $\sigma^2$ are one of the main distinguishing features between LQC and GR. In GR, the unboundedness of these quantities results in break down of geodesic evolution, a characetristic feature of singularities. In this work, we have established for the first time, that such divergences do not occur in the loop quantization of Bianchi-II and Bianchi-IX spacetimes in LQC. It has been shown for isotropic models, that these bounds play an important role for generic resolution of singularities in isotropic spacetimes in LQC \cite{ps09}. Novel results obtained in this work, take us a step closer to prove generic resolution of singularities in Bianchi-II and Bianchi-IX spacetimes.

 Given the remarkable coherence of results for the isotropic and Bianchi-I models, and a promising strategy to quantize Bianchi-II and Bianchi-IX models which is consistent with the loop quantization of the Bianchi-I model, one may have expected that  answers to the above questions would be on the lines of the results obtained in the previous models. However, we show that this expectation turns out to be not true and the answers leads to some surprises. We demonstrate that in the
Bianchi-II model, the energy density has a global maxima. The expansion and shear scalars are bounded  only if one imposes energy conditions on the matter content. In particular, it is important that the energy density has a lower bound (a simple requirement) which yields an upper bound on the shear scalar. The bounds are not universal and their values depend on the imposed energy conditions. Recall that for the isotropic and Bianchi-I models, the bounds on $\rho$, $\theta$ and $\sigma^2$ turned out to be universal and did not depend on energy conditions. It is to be noted that the in the effective Hamiltonian, matter Hamiltonian is treated as if Fock quantized. If matter is polymer quantized, in the way geometry is, results on dependence on energy conditions can a priori change. For the Bianchi-IX model, the inverse triad corrections are critical to obtain a bound on the energy density and the expansion and shear scalars when the approach to singularity in the classical theory is isotropic. In the Bianchi-IX model, energy conditions are shown to play little role. Inverse triad corrections are also important in other types of singularities, such as the cigar and pancake singularities, to achieve a finite value of the above quantities. In the effective description, singularity is resolved in the sense that the energy density, and the expansion and shear scalars  are finite. However, there also exist following two mathematical possibilities. If the effective dynamics allows solutions where two of the triads tend to zero and the third diverges, then $\rho_q$, $\theta$ and $\sigma^2$ diverge in Bianchi-IX spacetime.
Also,  if the physical solutions exist such that one of the triads vanish and the other two diverge,  then $\theta$ and $\sigma^2$ diverge. We emphasize, that due to the underlying assumptions in the derivation of effective Hamiltonian constraint, it is  not clear whether the effective dynamics description is valid when above two cases arise.

Though the bounds on energy density, and the expansion and shear scalars for Bianchi-II and Bianchi-IX models turn out to be different from the Bianchi-I case, it is important to emphasize that these bounds are not optimal, as shown by a separate analysis \cite{gs2}.
Nevertheless, results on Bianchi-II and Bianchi-IX models stand in sharp contrast to those obtained for the isotropic and Bianchi-I models, where for the first time the role of energy conditions  and the inverse volume modifications become important.\footnote{The inverse triad modifications lead to some interesting effects due to the choice of lapse in the Bianchi-IX model. These issues will be reported separately \cite{gs3}.}   Since the only change in the quantization strategy of these models, in comparison to the earlier ones, is in the way field strength is expressed in terms of connection, the cause of above differences lies in the usage of the connection operator approach. This quantization strategy is consistent with the
holonomy based quantization of the Bianchi-I model \cite{awe2}, but important differences can arise in presence of spatial curvature, as is evident from the analysis of Bianchi-IX model. The effective Hamiltonian of the loop quantization of Bianchi-IX spacetime does not lead to the one for the holonomy based loop quantization of $k=1$ model in the isotropic limit \cite{apsv}. Rather, its isotropic limit corresponds to the connection operator quantization of isotropic spatially closed model \cite{ck}, where the expansion scalar does not have a maxima. We also find that connection operator approach for $k=-1$ model suffers from the same problem as shown in Appendix B. The unboundedness of $\theta$ in the connection operator approach for the isotropic $k=\pm 1$ models,  does not directly affect its viability in the anisotropic models. However, these results show that this approach leads to a qualitatively different behavior, in comparison to the holonomy based quantization, in the presence of spatial curvature.

We conclude with a discussion of some open questions resulting from our analysis. In this work, we have used the effective Hamiltonian framework for the entire range of volume.  In previous studies, extensive numerical simulations have confirmed the validity of effective spacetime description in isotropic models for universes where bounce occurs at volumes greater than Planck volume \cite{aps1,aps2,aps3,apsv,bp,ap} and also Bianchi-I models (albeit for a different quantization) \cite{szulc_b1,madrid_b1}. It will be reasonable to expect the validity of bounds for the scales of interest, since  the maximum values of physical quantities occur at volumes greater than the Planck volume.    Nevertheless, this is an important issue which deserves a careful examination in future. In particular to understand the behavior of energy density, expansion and shear scalars when approach to classical singularity is non-isotropic in the Bianchi-IX model. For this one has to derive effective Hamiltonian dynamics using embedding approach in the presence of anisotropies and examine its validity at small volumes. The pertinent question is whether such an analysis yields further corrections to the effective Hamiltonian constraint, and if so, how are the bounds affected in Bianchi-II and Bianchi-IX models. It will be interesting to see if such modifications bring the maximum values of physical quantities closer to the isotropic and Bianchi-I models. The second issue concerns with the energy conditions. In GR, these conditions play an important role in the proof of singularity theorems. Their role in LQC has so far been irrelevant. Analysis of modified Einstein's equations in Bianchi-II model reveal, that in LQC, expansion and shear scalars are unbounded if the energy density is unbounded below. This suggests that for cases where the latter is satisfied, singularities may not be resolved by quantum geometry effects in LQC. This feature confirms with the general expectations that quantum gravity may not resolve all the singularities and weed out the unphysical situations with arbitrary negative energy, as considered above \cite{horo_myers}. In future work, it will be important to investigate this issue in detail and classify the singularities corresponding to such solutions. Our analysis also shows the non-trivial role played by the inverse volume modifications in resolution of isotropic singularity in the case when spatially topology is compact. Perhaps, due to simplicity of the model, this was never revealed in the isotropic case.\footnote{Novel phenomenological implications have though been discussed, see for eg. \cite{joao_ps}.}  However, for Bianchi-IX model it is the interplay of both the non-local field strength tensor and the inverse volume modifications which results in bounds on physical quantities. These results seem to point out that in general for spatially compact manifolds, physics of singularity resolution might be incomplete without inverse volume modifications. Finally, it is pertinent to revisit the connection operator approach, by demanding that the quantization of  Bianchi-IX model be consistent with holonomy based quantization of $k=1$ model. This is important as it will establish a consistency between the connection operator approach and holonomy based quantization in the presence of spatial curvature. It will be interesting to probe  the way results obtained in this work  are affected in such a quantization.

\vskip0.5cm

\acknowledgments
We thank A. Ashtekar, A. Corichi, J. Diaz-Polo, T. Pawlowski, J. Pullin and E. Wilson-Ewing for useful discussions. We are grateful to A. Ashtekar, A. Corichi, A. Karami, E. Montoya and E. Wilson-Ewing for a careful reading of the previous draft and comments which led to its significant improvement. This work is supported by NSF grant PHY1068743.

\appendix

\section{Expansion scalar in  $k=1$ model}
In this appendix, we summarize the derivation of the expansion scalar using the effective Hamiltonian constraint of $k=1$ model in LQC. Based on the techniques used for several models, to express field strength in terms of holonomies over a closed loop, the quantization of this model was performed in Ref. \cite{apsv}, which showed absence of singularity and a non-singular bounce when $\rho = \rmax \approx 0.41 \rho_{\mathrm{Pl}}$. More recently, an alternative quantization of this spacetime has been proposed \cite{ck}, which is based on the connection operator approach used for Bianchi-II and Bianchi-IX models \cite{awe2,we1}. Using the terminology from Ref. \cite{ck}, we will refer to these  as ``holonomy based'' and ``connection operator'' quantizations respectively.

The spatial manifold in $k=1$ model is $\mathbb{S}^3$ with radius $\ell_o = V_o^{1/3}$, where the fiducial volume $V_o = 16 \pi^2$.
In terms of the connection $c$ and triad $p$ variables, satisfying $\{c,p\} = 8 \pi G \gamma/3$,  the classical Hamiltonian constraint is given by
\be
\label{k1ham}\mathcal{H}_{\mathrm{cl}}=-\frac{3{p}^2}{8 \pi G\gamma^{2}}\left[\left(c-\varepsilon \f{\ell_o}{2}\right)^2+\f{\ell_o^2}{4}\gamma^2\right]+\mathcal{H}_{\rm matt} V
\ee
where $\varepsilon$ denotes the orientation of the triad.\footnote{Unlike the main body of this work, we will allow both the orientations of the triads in the appendices. This is to facilitate a comparison between the isotropic limits of the Bianchi model with the $k=1$ model. We thank E. Wilson-Ewing for a discussion on this issue in $k=1$ model.} The triad $p$ is related to the scale factor $a(t)$ in the spacetime metric
\be
ds^2=-dt^2+ a^2(t)\left[\frac{dr^2}{1-r^2}+r^2\left(d\theta^2+\sin^2\phi d\phi^2 \right)\right]
\ee
as $p=\varepsilon a^2 \ell_o^2$. The equation of motion for the triad can be derived using the Hamilton's equations:
\be
\label{pdot} \dot{p}=\f{2\sqrt{|p|}}{\gamma}\left(c-\f{\varepsilon \ell_o}{2}\right)
\ee
where the `dot' refers to the derivative with respect to the proper time. From which one obtains the relation
\be
c = \varepsilon \ell_o \left(\gamma \dot a + \f{1}{2} \right) ~.
\ee

Using these equations, it is straightforward to obtain the expansion scalar for the comoving observers: $\theta = 3 H = 3 \dot a/a$, which turns out to be
\be
	 \theta=3\frac{\left(c-\varepsilon\f{\ell_o}{2}\right)}{\gamma \sqrt{|p|}} ~.
\ee
In the classical theory, the approach to  singularity is characterized by a divergence in the connection $c$, along with a divergence in $1/{\sqrt{|p|}}$, causing the expansion scalar to become infinite. Let us now discuss the way this behavior changes in LQC for the holonomy based quantization performed in Ref. \cite{apsv} and the connection based quantization \cite{ck}.

\vskip0.5cm
{\it{Holonomy based quantization}}: For the loop quantization of $k=1$ model, proposed in Ref. \cite{apsv}, the effective Hamiltonian constraint for  lapse $N = V$ is given by
\be
\label{k1hapsv}	\mathcal{H}_{\rm eff}=-\frac{3}{8 \pi G\gamma^2}\frac{p^2}{\bar{\mu}^2}\left[\sin^2\left(\bar{\mu}\left(c-\varepsilon\f{\ell_o}{2}\right)\right)-\chi\right]+\mathcal{H}_{\rm matt} V
\ee
where $\chi=\sin^2\left({\bar{\mu}\varepsilon\f{l_0}{2}}\right)-(1+\gamma^2)(\bar{\mu}\varepsilon\f{l_0}{2})^2$, $\bar \mu^2 = \lambda^2/p$ and $\lambda^2=4\sqrt{3}\pi\gamma{l_p}^2$. This includes the inverse triad modifications, which appear only in the matter Hamiltonian.
The Hamilton's equation for triad yields
\be
\dot{p}=\frac{2p}{\gamma \lambda}\sin\left({\bar{\mu}\left(c-\varepsilon\f{\ell_o}{2}\right)}\right)\cos\left({\bar{\mu}\left(c-\varepsilon\f{\ell_o}{2}\right)}\right) ~,
\ee
and the expansion parameter turns out to be
\be
\label{rightexp}\theta=\frac{3}{\gamma \lambda}\sin\left({\bar{\mu}\left(c-\varepsilon\f{\ell_o}{2}\right)}\right)\cos\left({\bar{\mu}(c-\varepsilon\f{\ell_o}{2})}\right) ~.
\ee
This is a bounded function with a  maximum value given by
\be\label{thetamaxk1}
\theta_{\rm max}=\f{3}{2\gamma \lambda} ~,
\ee
which agrees with the maximum value in isotropic \cite{ps09} and Bianchi-I model \cite{cs09}.
It is important to note that the above value is not affected by the inverse triad modifications, or the choice of energy conditions i.e. the specific form of matter Hamiltonian. In this sense, the maximum value of the expansion scalar is universal.  \\

{\it{Connection operator quantization}}: Based on the loop quantization of Bianchi-II and Bianchi-IX spacetimes, one can quantize the $k=1$ model using an alternative method which avoids expressing field strength in terms of holonomies over a closed loop. Instead, the field strength is expressed in terms of the connection operator obtained from holonomies over open edges. This quantization results in the following effective Hamiltonian: \footnote{Effective dynamics for this model in the absence of the inverse triad corrections has been studied in Ref. \cite{ck}, however problems with expansion scalar as found here were not discussed.}
\be \label{heff_k1conn}
	\mathcal{H}_{\rm eff}=-\frac{3p^2}{8 \pi G\gamma^2} \left[\left(\frac{\sin(\bar{\mu}c)}{\bar{\mu}}-\varepsilon\f{\ell_o}{2}\right)^2+\f{\ell_o^2\gamma^2}{4} \right]+\mathcal{H}_{\rm matt} V ~.
\ee
The Hamilton's equation for the triad yields
\be
\dot{p}=\frac{\sqrt{|p|}}{\gamma}\left[\frac{\sin\left(2\bar{\mu}c\right)}{\bar{\mu}}-2\varepsilon\f{\ell_o}{2}\cos\left(\bar{\mu}c\right)\right] ~,
\ee
and the expansion parameter turns out to be
\be\label{theta_connk1}
\theta=\frac{3}{2\gamma}\left[\frac{1}{\lambda}\sin\left({2\bar{\mu}c}\right)-\varepsilon\f{\ell_o}{2}\frac{2}{\sqrt{|p|}}\cos\left({\bar{\mu}c}\right)\right] ~.
\ee
If one assumes the validity of ${\cal H}_{\rm eff}$ for all values of triad, the expansion scalar turns out to be unbounded: as $p\rightarrow 0$, $\theta \rightarrow -\infty$.\footnote{It is possible that the effective Hamiltonian description may attain additional corrections as $p\rightarrow0$ which may regularize the above unbounded behavior. At the present stage, this is an open issue.} This behavior is in agreement with that of the expansion scalar of the loop quantization of Bianchi-IX model {\it without} the inverse triad corrections.

In conclusion, we find that there is an important qualitative difference in the behavior of expansion scalar in the  holonomy based quantization (eq.(\ref{rightexp})) and the connection operator  quantization (eq.(\ref{theta_connk1})) for the $k=1$ model. The expansion scalar in the holonomy based quantization turns out to be bounded by a universal value irrespective of the choice of matter, where as for the connection operator quantization it is divergent. Since the latter is tied to the existence of singularities, this result shows that with in the applicability of effective Hamiltonian constraint, connection operator quantization for $k=1$ model has a drawback.

\section{Expansion scalar for $k=-1$ model in LQC}
Loop quantization of $k=-1$ model has been proposed in Refs. \cite{kv,szulc}. In this model, the Ashtekar-Barbero connection contains off-diagonal terms which poses technical problems in construction of the quantum theory using holonomies as elementary variables. In order to overcome these issues,  loop quantization has been performed by considering holonomies of the extrinsic curvature. The quantum Hamiltonian constraint turns out to be non-singular with features similar to the quantization of other isotropic models \cite{kv,szulc}. The resulting physics has been analyzed using the effective Hamiltonian constraint which reveals resolution of strong singularities \cite{psvt}. A drawback of this quantization is that field strength is not expressed in terms of the holonomies of the connection\footnote{Nevertheless, it has been argued that in the symmetry reduced setting, due to gauge fixing it is possible to consider extrinsic curvature at the same footing as the connection \cite{kv}.}, which motivates to consider alternate quantizations of $k=-1$ model in LQC. A possible avenue would be to quantize it on the lines of Bianchi-II and Bianchi-IX models using connection operator approach which would bypass the above problem \cite{aa_private}. Here we compute the expansion scalars in both of these approaches using the effective Hamiltonian constraint. As in the $k=1$ model, we will refer to them as ``holonomy based'' and ``connection operator'' quantizations respectively.

The classical Hamiltonian constraint for the $k=-1$ model (in lapse $N = V$) is given by
\be\label{hclkmin1}
\mathcal{H}_{\rm cl}=-\frac{3p^2}{8 \pi G\gamma^{2}}\left[\left(c+\varepsilon \f{l_o}{2}\right)^2-\f{l_o^2}{4}\gamma^2\right]+\mathcal{H}_{\rm matt} V~
\ee
where $l_o$ (not to be confused with $\ell_o$ introduced earlier) refers to the cube root of the fiducial volume $V_o$ of the fiducial cell ${\cal V}$ needed to introduce the symplectic structure on the non-compact manifold. The triad $p$ is related to scale factor as
\be
p=\varepsilon a^2 l_o^2,
\ee
and its time variation is given by
\be
\label{pdot-1} \dot{p}=\f{2\sqrt{|p|}}{\gamma}\left(c+\f{\varepsilon l_o}{2}\right) ~.
\ee
Substituting the expression for $p$ in terms of scale factors into the eq. \ref{pdot-1} we obtain,
\be
c+\f{\varepsilon l_o}{2}=\gamma \varepsilon \dot{a} l_o ~,
\ee
using which one finds the expansion scalar as:
\be
\theta=3\frac{\left(c+\varepsilon\f{l_o}{2}\right)}{\gamma \sqrt{|p|}} ~.
\ee
From the above expression, we find that the expansion scalar in the classical $k=-1$ model diverges as the singularity is approached.

\vskip0.5cm
{\it Holonomy based quantization:} The effective Hamiltonian constraint for the loop quantization of $k=-1$ model as performed in Refs. \cite{kv,szulc} is given by
\be
\label{righthamilt-1}	\mathcal{H}_{\rm eff}=-\frac{3}{8 \pi G\gamma^2}\frac{p^2}{\bar{\mu}^2}\left[\sin^2\left(\bar{\mu}\left(c+\varepsilon\f{l_o}{2}\right)\right)+\chi\right]+\mathcal{H}_{\rm matt} V
\ee
where $\chi=-\gamma^2(\bar{\mu}\varepsilon\f{l_0}{2})^2$ and $\bar \mu$ is the same as in $k=1$ model. Since the spatial manifold is noncompact, there are no inverse triad corrections in $\mathcal{H}_{\rm eff}$. A straightforward calculation, as performed for the $k=1$ model shows that the expansion scalar turns out to be
\be
\label{rightexp-1}\theta=\frac{3}{\gamma \lambda}\sin\left({\bar{\mu}\left(c+\varepsilon\f{l_o}{2}\right)}\right)\cos\left({\bar{\mu}\left(c+\varepsilon\f{l_o}{2}\right)}\right)
\ee
where we have used the Hamilton's equation for the triad:
\be
\dot{p}=\frac{2p}{\gamma \lambda}\sin\left({\bar{\mu}\left(c+\varepsilon\f{l_o}{2}\right)}\right)\cos\left({\bar{\mu}\left(c+\varepsilon\f{l_o}{2}\right)}\right) ~.
\ee
The expansion scalar for the holonomy based quantization of $k=-1$ model thus turns out to be a bounded function with the same maximum value as the expansion scalar for $k=1$ model (eq.(\ref{thetamaxk1})).\\

{\it Connection operator quantization:} The effective Hamiltonian constraint in this case can be obtained by replacing $c$ with ${\sin(\bar{\mu}c)}/{\bar{\mu}}$ in the classical Hamiltonian constraint (\ref{hclkmin1}):
\be
	\mathcal{H}_{\rm eff}=-\frac{3}{8 \pi G\gamma^2} p^2\left[\left(\frac{\sin(\bar{\mu}c)}{\bar{\mu}}+\varepsilon\f{l_o}{2}\right)^2-\f{l^2_o\gamma^2}{4} \right]+\mathcal{H}_{\rm matt} V ~.
\ee
The resulting Hamilton's equation for the triad becomes
\be	 \dot{p}=\frac{\sqrt{|p|}}{\gamma}\left[\f{\sin\left(2\bar{\mu}c\right)}{\bar{\mu}}+2\varepsilon\f{l_o}{2}\cos\left(\bar{\mu}c\right)\right]
\ee
which leads to the following expression for the expansion parameter,
\be
\theta=\frac{3}{2\gamma}\left[\frac{1}{\lambda}\sin\left({2\bar{\mu}c}\right)+\varepsilon\f{l_o}{2}\frac{2}{\sqrt{|p|}}\cos\left({\bar{\mu}c}\right)\right] ~.
\ee
Unlike the behavior of the expansion scalar in the holonomy based quantization, $\theta$ does not have a maxima.  The connection operator approach encounters the same limitation as the $k=1$ model regarding the behavior of the expansion scalar. Thus, there are important qualitative differences in the resulting physics from holonomy based and connection operator quantizations in LQC.

\end{document}